\documentclass[prd,aps,superscriptaddress,twocolumn,floatfix,nofootinbib]{revtex4}
\pdfoutput=1

\usepackage{amsfonts}
\usepackage{amsmath}
\usepackage{amssymb}
\usepackage{bm}
\usepackage{dcolumn}
\usepackage{graphicx}   
\usepackage[latin1]{inputenc}
\usepackage{latexsym}
\usepackage{rotating}
\usepackage{hyperref}
\usepackage{graphicx}
\usepackage{color}

\newcommand\be{\begin{equation}}
\newcommand\ba{\begin{eqnarray}}
\newcommand\ee{\end{equation}}
\newcommand\ea{\end{eqnarray}}

\newcommand{\cn}{\operatorname{cn}}
\newcommand{\dd}{\mathrm{d}}

\begin{document}

\title{Indirect Parametric Resonance of the Electromagnetic Field Driven by an Oscillating \(SU(2)\) Dark Matter Condensate}

\author{Tatsuya Daniel}
\email{tdaniel@physics.mcgill.ca}
\affiliation{Department of Physics, McGill University, Montr\'{e}al,
  QC, H3A 2T8, Canada}
\affiliation{Trottier Space Institute, Department of Physics, McGill
University, Montr\'{e}al, QC, H3A 2T8, Canada}
  
\author{Vahid Kamali}
\email{vahid.kamali2@mcgill.ca}
\affiliation{Department of Physics, McGill University, Montr\'{e}al,
  QC, H3A 2T8, Canada}
\affiliation{Trottier Space Institute, Department of Physics, McGill
University, Montr\'{e}al, QC, H3A 2T8, Canada}
 \affiliation{School of Continuing Studies, McGill University, Montr\'{e}al,
QC, H3A 2T5, Canada} 
\affiliation{Department of Physics, Bu-Ali Sina (Avicenna) University,
Hamedan 65178, 016016, Iran}

\author{Robert Brandenberger}
\email{rhb@physics.mcgill.ca}
\affiliation{Department of Physics, McGill University, Montr\'{e}al,
  QC, H3A 2T8, Canada}
\affiliation{Trottier Space Institute, Department of Physics, McGill
University, Montr\'{e}al, QC, H3A 2T8, Canada}
\affiliation{Kavli IPMU (WPI), UTIAS, The University of Tokyo, 
5-1-5 Kashiwanoha, Kashiwa, Chiba 277-8583, Japan}


\begin{abstract}

We study a local patch of an axion-like dark sector in Minkowski spacetime, containing an initially homogeneous and isotropic non-Abelian \(SU(2)\) condensate, and a real pseudoscalar field \(\chi\),  coupled to an Abelian \(U(1)\) gauge field which could be that of usual electromagnetism. The pseudoscalar couples directly to both gauge sectors through Chern-Simons interactions, while the \(U(1)\) field couples to the \(SU(2)\) condensate only indirectly, through the pseudoscalar. We show analytically that a homogeneous oscillating \(SU(2)\) background \(Q(t)\) acts as a periodic source for \(\chi\), generating a homogeneous oscillatory condensate \(\dot\chi\) that in turn modulates the frequency of the \(U(1)\) helicity modes. In the linear regime this produces a Hill equation, and when the first harmonic dominates it reduces to a Mathieu equation. We derive the approximate resonance conditions, the leading Floquet exponent, and the conditions for a two-stage enhancement \(Q \to \chi \to U(1)\). We also highlight an important point: in the exactly periodic, zero-bias limit, the indirect resonance is not intrinsically chiral, because the two Abelian helicities are related by a half-period time shift.

\end{abstract}

\maketitle

\section{Motivation} 
\label{sec:intro}

It has recently been pointed out \cite{BFJ} that an oscillating ultralight pseudoscalar dark matter field coherently oscillating on cosmological scales and coupled via a Chern-Simons-like interaction term to electromagnetism\footnote{See \cite{original} for early works studying this interaction.} induces an infrared instability of the electromagnetic gauge field modes and is able to generate magnetic fields of sufficient strength to explain the blazar observations \cite{blazar}.  Depending on the model parameters,  either a tachyonic instability channel \cite{BFJ} or a narrow band parametric resonance channel \cite{Nirmalya} dominates.  In \cite{Kamali}, it was shown that the basic mechanism also can be realized if the ultralight dark matter field is a scalar condensate.  Through the coupling of the induced cosmological magnetic field to dark matter fluctuations, a sufficient flux of Lyman-Werner photons can be generated to open up the direct collapse black hole formation window \cite{Ashu} (see also \cite{JKB}).

In this work, we ask the question whether an oscillating vector field dark matter condensate can also lead to the generation of magnetic fields.  In light of the role which vector field condensates can play in cosmology (e.g. in the context of inflation \cite{Azadeh}), there is interest in studying this question.  As realized in \cite{Azadeh}, it is - unlike in the case of a $U(1)$ gauge field - possible to generate a homogeneous and isotropic condensate if the gauge group is $SU(2)$ (or contains a $SU(2)$ factor).  However, because of gauge invariance, it is difficult to generate a direct coupling between the two gauge fields. It is more natural to envision the presence of an axionic field $\chi$ which couples to both $F_{\mu \nu}$ and $X_{\mu \nu}$, where $F$ and $X$ are the field strength tensors of the non-Abelian dark matter vector field $A_{\mu}$ and of the gauge field $C_{\mu}$ of electromagnetism,  respectively. In this case, the energy transfer between the oscillating condensate and the electromagnetic field takes place in two stages: in the first, an oscillating $\chi$ field condensate is excited by the $A_{\mu}$ field, and in the second, the oscillating $\chi$ field induces the generation of electromagnetic excitations.

More generally, we are interested in the indirect effect of an oscillating \(SU(2)\) condensate on an Abelian \(U(1)\) gauge field in a dark matter halo. The natural hierarchy is
\begin{equation}
SU(2)\ \text{condensate } Q(t)~\longrightarrow~\chi(t)~\longrightarrow~U(1)\ \text{modes}.
\end{equation}
The point is to isolate the induced periodic driving of the \(U(1)\) field.   We assume, as is done conventionally in discussions of light dark matter such as the axion \cite{axion}, that the initial coherent oscillations of the $SU(2)$ field are set up by some misalignment mechanism. For a recent proposal of such a mechanism, see \cite{misalign}. 

We work in a local patch in which the $SU(2)$ condensate can be viewed as homogeneous.  We can neglect cosmological expansion and approximate spacetime as Minkowski, with the metric tensor
\begin{equation}
\eta_{\mu\nu} = \mathrm{diag}(-1,+1,+1,+1) \, .
\end{equation}
This is a self-consistent approximation as long as the oscillation and induced instability time scales are much shorter than the gravitational time scale of the local patch (e.g.  the Hubble expansion time scale), and the relevant length scales are much smaller than the halo size. Cosmic expansion is then negligible, and all Hubble friction terms in the field equations of motion can be dropped.

\section{Action and Field Equations}

\noindent
We start from the flat-space action
\ba
S \, &=& \,  \int \dd^4x\, \Bigg[
-\frac{1}{2}\,\partial_\mu \chi \, \partial^\mu \chi
-U(\chi)
-\frac{1}{4}F^a_{\mu\nu}F^{a\mu\nu} \\
& & +\frac{\lambda_1}{4f}\,\chi\,F^a_{\mu\nu}\tilde F^{a\mu\nu}
-\frac{1}{4}X_{\mu\nu}X^{\mu\nu}
+\frac{\lambda_2}{4f}\,\chi\,X_{\mu\nu}\tilde X^{\mu\nu}
\Bigg], \nonumber
\label{eq:action}
\ea
where
\ba
F^a_{\mu\nu} &=& \partial_\mu A^a_\nu - \partial_\nu A^a_\mu - g_A \epsilon^{abc} A^b_\mu A^c_\nu,
\nonumber \\
X_{\mu\nu} &=& \partial_\mu C_\nu - \partial_\nu C_\mu,
\ea
and the dual tensors are defined by
\ba
\tilde F^{a\mu\nu} &=& \frac{1}{2}\epsilon^{\mu\nu\rho\sigma}F^a_{\rho\sigma},
\\
\tilde X^{\mu\nu} &=& \frac{1}{2}\epsilon^{\mu\nu\rho\sigma}X_{\rho\sigma},
\qquad
\epsilon^{0123}=+1. \nonumber
\ea

\noindent
Varying the action gives
\begin{align}
\Box\chi - U_{,\chi}
&= -\frac{\lambda_1}{4f}F^a_{\mu\nu}\tilde F^{a\mu\nu} - \frac{\lambda_2}{4f}X_{\mu\nu}\tilde X^{\mu\nu},
\label{eq:chi_general}
\\
D_\mu F^{a\mu\nu}
&= -\frac{\lambda_1}{f}(\partial_\mu\chi)\tilde F^{a\mu\nu},
\label{eq:su2_general}
\\
\partial_\mu X^{\mu\nu}
&= -\frac{\lambda_2}{f}(\partial_\mu\chi)\tilde X^{\mu\nu},
\label{eq:u1_general}
\end{align}
with \(\Box = -\partial_t^2 + \nabla^2\).

If we start out with initial conditions in which the energy resides in an oscillating $SU(2)$ condensate, then in the first step of the energy transfer, the right hand side of (\ref{eq:chi_general}) leads to the generation of an oscillating $\chi$ condensate. Inserting this condensate into the right hand side of (\ref{eq:u1_general}) induces an effective oscillating mass term for the $U(1)$ gauge field, which then via parametric resonance leads to an efficient magnetic field production.  For a low effective mass scale of the $SU(2)$ condensate, the resonance leads to the production of $U(1)$ gauge fields on large (cosmological) scales.  

Once a substantial fraction of the energy has flowed into the $\chi$ and $C_{\mu}$ fields,  backreaction terms (e.g. the right hand side of (\ref{eq:su2_general})) become important and destroy the coherence of the fields. This shuts off the resonance. In this paper, we will not consider backreaction effects (for an initial attempt at analyzing backreaction effects in the context of the model of \cite{BFJ}, see \cite{Nirmalya}). 

\section{Isotropic \(SU(2)\) Condensate and Reduced Background System}

To keep the non-Abelian background isotropic, we use the standard ansatz \cite{Azadeh}
\begin{equation}
A_0^a = 0,
\qquad
A_i^a = \delta_i^{\,a} Q(t),
\label{eq:su2_ansatz}
\end{equation}
where \(Q(t)\) is a single real amplitude. This gives
\begin{equation}
F^a_{0i} = \delta_i^{\,a}\dot Q,
\qquad
F^a_{ij} = -g_A \epsilon^{aij}Q^2,
\end{equation}
and therefore
\begin{equation}
F^a_{\mu\nu}\tilde F^{a\mu\nu} = -12 g_A Q^2 \dot Q.
\label{eq:FFdual}
\end{equation}

\noindent
We define the Abelian electric and magnetic fields by
\ba
E_{X,i} &=& X_{0i},
\qquad
B_{X,i} = \frac{1}{2}\epsilon_{ijk}X_{jk}, \nonumber \\
X_{\mu\nu}\tilde X^{\mu\nu} &=&  -4\,\mathbf E_X\cdot \mathbf B_X.
\label{eq:XXdual}
\ea

We start both the $\chi$ and the $C_{\mu}$ fields unexcited (modulo their quantum vacuum fluctuations).  The homogeneous $Q$ condensate will induce a homogeneous component $\chi(t)$ which grows in time.  For this component, Eq.~\eqref{eq:chi_general} reduces to
\be
\ddot\chi + U_{,\chi}
= -\frac{3g_A\lambda_1}{f}Q^2\dot Q - \frac{\lambda_2}{f}\,\mathbf E_X\cdot\mathbf B_X \, .
\label{eq:chi_red}
\ee
The equation for the condensate $Q(t)$ is (see (\ref{eq:su2_general}))
\be
\ddot Q + 2g_A^2Q^3
 = \frac{g_A\lambda_1}{f}\dot\chi\,Q^2 \, ,
\label{eq:Q_red}
\ee
 where the right hand side of the equation describes how the induced $\chi(t)$ leads to a backreaction effect on the condensate $Q(t)$. 
 
 The gauge field equation is
 \be
\ddot{\mathbf C} - \nabla^2\mathbf C - \frac{\lambda_2}{f}\dot\chi\,\nabla\times\mathbf C
= 0 \, ,
\label{eq:C_vector}
\ee
with initial conditions given by the quantum vacuum fluctuations.  The last term in this equation leads to the parametric resonant excitation of Fourier modes of $C$.

As long as fluctuations in $A_{\mu}$ and $\chi$ are negligible,  the  local energy density is
\begin{equation}
\rho_{\rm loc}
= \frac{1}{2}\dot\chi^2 + U(\chi)
+ \frac{3}{2}\dot Q^2 + \frac{3}{2}g_A^2Q^4
+ \frac{1}{2}\left(E_X^2 + B_X^2\right).
\end{equation}

\section{Evolution of the $SU(2)$ Condensate}

To isolate the indirect effect \(Q \to \chi \to U(1)\), we now work in the linear stage before significant backreaction from the Abelian sector. Concretely, we assume
\ba
\left|\frac{\lambda_2}{f}\mathbf E_X\cdot\mathbf B_X\right| &\ll& \left|\frac{3g_A\lambda_1}{f}Q^2\dot Q\right|, \nonumber \\
\left|\frac{g_A\lambda_1}{f}\dot\chi\,Q^2\right| &\ll& \left|2g_A^2Q^3\right|.
\label{eq:linear_regime}
\ea
Then we solve the system sequentially:
\begin{enumerate}
\item Solve for \(Q(t)\) to leading order without backreaction terms.
\item Use \(Q(t)\) as a source for \(\chi(t)\).
\item Use the induced \(\dot\chi(t)\) as the pump field for the \(U(1)\) helicity modes.
\end{enumerate}
 
\noindent
At leading order, Eq.~\eqref{eq:Q_red} becomes
\begin{equation}
\ddot Q + 2g_A^2Q^3 = 0.
\label{eq:Q_uncoupled}
\end{equation}
This equation has periodic solutions.  The relevant exact solution is
\begin{equation}
Q(t) = Q_0\,\cn\!\left(\sqrt{2}\,g_AQ_0 t + \delta,\,\frac{1}{2}\right),
\label{eq:Q_exact}
\end{equation}
where \(Q_0\) is the oscillation amplitude, \(\delta\) is a phase, and $\cn(u,m)$ denotes the Jacobi elliptic cosine function with modulus $m$. The period is
\begin{equation}
T_Q = \frac{4K(1/2)}{\sqrt{2}\,g_AQ_0},
\end{equation}
and the associated frequency is
\begin{equation}
\Omega_Q \equiv \frac{2\pi}{T_Q} = \frac{\pi\sqrt{2}\,g_AQ_0}{2K(1/2)} \simeq 1.20\, g_AQ_0 \, ,
\label{eq:OmegaQ_exact}
\end{equation}
where \(K(m)\) is the complete elliptic integral of the first kind.

This solution for $Q(t)$ can be expanded in temporal Fourier modes. For analytic estimates, it is sufficient to retain only the dominant Fourier harmonic and write
\begin{equation}
Q(t) \simeq Q_0\cos(\Omega_Q t).
\label{eq:Q_cosine}
\end{equation}
A more general odd-harmonic expansion can be written as
\begin{equation}
Q(t) = \sum_{r=0}^{\infty} Q_{2r+1}\cos\!\big[(2r+1)\Omega_Q t\big].
\label{eq:Q_fourier}
\end{equation}
Because the oscillation is centered around zero, only odd harmonics appear.

\section{Scalar Field Response Induced by the \(SU(2)\) Condensate}

We assume that the pseudoscalar field $\chi$ starts at the minimum of its potential. Expanding about the minimum, we can consider the potential $U(\chi)$ to be quadratic (to leading order), i.e.
\begin{equation}
U(\chi) = \frac{1}{2}m_\chi^2\chi^2 \, .
\end{equation}
To leading order, the homogeneous condensate can be decomposed as
\begin{equation}
\chi(t) = \chi_{\rm dm}(t) + \delta\chi_Q(t),
\label{eq:chi_split}
\end{equation}
where \(\chi_{\rm dm}\) solves the homogeneous equation
\begin{equation}
\ddot\chi_{\rm dm} + m_\chi^2\chi_{\rm dm} =0,
\end{equation}
and \(\delta\chi_Q\) is the induced response sourced by the \(SU(2)\) background.~Neglecting the Abelian backreaction, Eq.~\eqref{eq:chi_red} gives
\begin{equation}
\ddot{\delta\chi}_Q + m_\chi^2\delta\chi_Q = -\frac{3g_A\lambda_1}{f}Q^2\dot Q.
\label{eq:deltachi_general}
\end{equation}

We focus on the periodic particular solution of this equation, which captures the steady-state response of the scalar field to the oscillating $SU(2)$ condensate.~Transient contributions from the homogeneous solution are neglected. Since we assume that \(\chi_{\rm dm}\) is initially fixed at the minimum of the potential, we have \(\chi_{\rm dm}(0)\) = \(\dot\chi_{\rm dm}(0)\) = 0. Therefore, we just consider the enhancement of the field sourced by $Q$. Given the time evolution of the source term $Q(t)$, we can solve for $\delta\chi_Q(t)$ by considering the Green's function method.

\subsection{General Fourier solution}

Using the expansion \eqref{eq:Q_fourier}, the source term can be decomposed as
\begin{equation}
-\frac{3g_A\lambda_1}{f}Q^2\dot Q
\equiv \sum_{r=0}^{\infty} S_{2r+1}\sin\!\big[(2r+1)\Omega_Q t\big],
\label{eq:S_series}
\end{equation}
again with odd harmonics only. Away from exact forced resonance, the particular solution is
\begin{equation}
\delta\chi_Q(t)
= \sum_{r=0}^{\infty}
\frac{S_{2r+1}}{m_\chi^2-(2r+1)^2\Omega_Q^2}
\sin\!\big[(2r+1)\Omega_Q t\big].
\label{eq:deltachi_series}
\end{equation}
Differentiating once gives
\begin{equation}
\dot{\delta\chi}_Q(t)
= \sum_{r=0}^{\infty}
\frac{(2r+1)\Omega_Q\,S_{2r+1}}{m_\chi^2-(2r+1)^2\Omega_Q^2}
\cos\!\big[(2r+1)\Omega_Q t\big].
\label{eq:deltachidot_series}
\end{equation}
This is the basic result: the oscillating \(SU(2)\) condensate creates an oscillatory \(\dot\chi\), and it is this quantity that drives the Abelian modes $C_k(t)$.

\subsection{Single-harmonic approximation}

\noindent
If we keep only the leading harmonic \eqref{eq:Q_cosine}, then
\begin{equation}
Q^2\dot Q = -\frac{\Omega_QQ_0^3}{4}\left[\sin(\Omega_Q t)+\sin(3\Omega_Q t)\right].
\end{equation}
Therefore, the source coefficients are
\begin{equation}
S_1 = S_3 = \frac{3g_A\lambda_1\Omega_QQ_0^3}{4f}.
\label{eq:S13}
\end{equation}
The induced scalar response is then
\begin{equation}
\delta\chi_Q(t)
\simeq \frac{S_1}{m_\chi^2-\Omega_Q^2}\sin(\Omega_Q t)
+ \frac{S_3}{m_\chi^2-9\Omega_Q^2}\sin(3\Omega_Q t),
\label{eq:deltachi_13}
\end{equation}
and
\begin{equation}
\dot{\delta\chi}_Q(t)
\simeq \frac{\Omega_QS_1}{m_\chi^2-\Omega_Q^2}\cos(\Omega_Q t)
+ \frac{3\Omega_QS_3}{m_\chi^2-9\Omega_Q^2}\cos(3\Omega_Q t).
\label{eq:deltachidot_13}
\end{equation}

A strong enhancement occurs when \(m_\chi\) is close to one of the odd harmonics,
\begin{equation}
m_\chi \approx (2r+1)\Omega_Q.
\label{eq:first_stage_resonance}
\end{equation}
Near exact resonance, the perturbative expressions above are replaced by a secular solution, for example
\begin{equation}
\delta\chi_Q(t) \sim \frac{S_n}{2m_\chi}\, t\cos(m_\chi t),
\qquad n\Omega_Q = m_\chi,
\label{eq:chi_secular}
\end{equation}
up to an irrelevant phase. This is the first stage of the cascade.

Away from a resonant point\footnote{The periodic analysis used here applies away from the exact resonance condition, where the response remains approximately bounded over the relevant time scales.}, we can estimate the fraction of the initial energy in the $SU(2)$ condensate which is transferred to $\chi$. For $m_{\chi} > \Omega_Q$, we obtain
\be \label{frac1}
\frac{\rho_{\chi}}{\rho_Q} \, \sim \, \lambda_1^2 \frac{1}{n^2} \, ,
\ee
while for $m_{\chi} < \Omega_Q$ we obtain
\be \label{frac2}
\frac{\rho_{\chi}}{\rho_Q} \, \sim \, n^2 \lambda_1^2 g_A^2 \bigg( \frac{Q_0}{f}\bigg)^2 \, .
\ee
From these results, we see that the fraction of the initial energy in $Q$ which is transferred to the oscillating $\chi$ condensate is model-dependent but can be quite large.  

\section{Indirect Pumping of the Abelian Helicity Modes}

We now insert the induced \(\dot\chi\) into the \(U(1)\) equation. In Coulomb gauge,
\begin{equation}
C_0 = 0,
\qquad
\nabla\cdot\mathbf C =0 \, .
\end{equation}
We expand the transverse Abelian field in Fourier modes:
\begin{equation}
\mathbf C(t,\mathbf x)
= \sum_{\lambda=\pm}\int \frac{\dd^3k}{(2\pi)^{3/2}}
\,\bm\epsilon_\lambda(\hat{\mathbf k})\,
\mathcal C_\lambda(t,k)
\,e^{i\mathbf k\cdot\mathbf x},
\end{equation}
with helicity vectors satisfying
\begin{equation}
i\mathbf k\times\bm\epsilon_\lambda = \lambda k\,\bm\epsilon_\lambda,
\qquad
\lambda=\pm 1.
\end{equation}
Then Eq.~\eqref{eq:C_vector} reduces mode-by-mode to
\ba
\ddot{\mathcal C}_\lambda &+& \omega_\lambda^2(t,k)\,\mathcal C_\lambda =0, \\
\omega_\lambda^2(t,k) &=& k^2 - \lambda k\frac{\lambda_2}{f}\dot\chi(t) \, . \nonumber
\label{eq:Clambda_general}
\ea
If we isolate only the \(SU(2)\)-induced piece \(\delta\chi_Q\), then
\begin{equation}
\ddot{\mathcal C}_\lambda
+ \left[k^2 - \lambda k\frac{\lambda_2}{f}\dot{\delta\chi}_Q(t)\right]\mathcal C_\lambda =0.
\label{eq:Clambda_indirect}
\end{equation}
Using the general series \eqref{eq:deltachidot_series}, we obtain the Hill equation
\begin{equation}
\ddot{\mathcal C}_\lambda
+ \left[k^2 - \lambda\sum_{r=0}^{\infty} k \beta_{2r+1}(k)
\cos\!\big[(2r+1)\Omega_Q t\big]\right]\mathcal C_\lambda =0,
\label{eq:Hill_general}
\end{equation}
with coefficients
\begin{equation}
\beta_{2r+1}(k)
\equiv \frac{\lambda_2}{f}
\frac{(2r+1)\Omega_Q\,S_{2r+1}}{m_\chi^2-(2r+1)^2\Omega_Q^2}.
\label{eq:beta_general}
\end{equation}
This equation demonstrates the indirect mechanism by which the $SU(2)$ dark sector vector field condensate can generate $U(1)$ gauge field modes via the intermediary $\chi$ field.

Considering the equation (\ref{eq:Hill_general}), we see that there are two possible resonance channels. For very small values of $k$, values for which the $k^2$ term is negligible, there is a tachyonic resonance channel like the one studied in \cite{BFJ}: during half of the oscillation period of the $\chi$ condensate, there is an effective negative $m^2$ term in the equation of motion, driving exponential growth, while in the other half period ${\mathcal C}_\lambda$ is oscillating.  If the rate $\mu_k$ of the exponential growth is larger than the frequency of oscillation, then the net growth of ${\mathcal C}_\lambda$ is exponential. The second resonance channel is narrow band parametric resonance, which occurs when the mode frequency $k$ is commensurate with the driving frequency, as will be studied below\footnote{Note that the physics involved here is analogous to the physics of preheating after inflation. See \cite{TB, DK, STB, KLS, tachyonic} for original references, \cite{ABCM, Karouby} for reviews, and \cite{TB2} for a brief historical review.}.  As we will show later, for parameter values appropriate for ultralight vector dark matter, the tachyonic decay channel is suppressed, and the narrow band parametric resonance channel dominates.

Keeping only the first and third harmonics from \eqref{eq:deltachidot_13}, the mode equation becomes
\begin{equation}
\ddot{\mathcal C}_\lambda
+ \Big[k^2 - k\lambda\beta_1(k)\cos(\Omega_Q t) - k\lambda\beta_3(k)\cos(3\Omega_Q t)\Big]\mathcal C_\lambda =0,
\label{eq:Clambda_13}
\end{equation}
where
\begin{align}
\beta_1(k)
&= \frac{3g_A\lambda_1\lambda_2}{4f^2}\,
\frac{\Omega_Q^2Q_0^3}{m_\chi^2-\Omega_Q^2},
\label{eq:beta1}
\\
\beta_3(k)
&= \frac{9g_A\lambda_1\lambda_2}{4f^2}\,
\frac{\Omega_Q^2Q_0^3}{m_\chi^2-9\Omega_Q^2}.
\label{eq:beta3}
\end{align}
Eq.~\eqref{eq:Clambda_13} is already sufficient to discuss the resonance using analytical arguments.

If the first harmonic dominates,
\begin{equation}
|\beta_1| \gg |\beta_3|,
\end{equation}
we may drop the third harmonic and,  in terms of the dimensionless time variable
\begin{equation}
z = \frac{\Omega_Q t}{2},
\end{equation}
Eq.~\eqref{eq:Clambda_13} reduces to Mathieu form,
\begin{equation}
\frac{\dd^2\mathcal C_\lambda}{\dd z^2}
+ \left[A_k - 2q_\lambda\cos(2z)\right]\mathcal C_\lambda =0,
\label{eq:Mathieu}
\end{equation}
with
\ba
A_k &=& \frac{4k^2}{\Omega_Q^2}, \\
q_\lambda &=& \lambda\frac{2k\beta_1(k)}{\Omega_Q^2}
= \lambda\frac{3g_A\lambda_1\lambda_2}{2f^2}\,
\frac{kQ_0^3}{m_\chi^2-\Omega_Q^2} \, . \nonumber
\label{eq:Akqk}
\ea

The tachyonic instability band holds for values of $k$ smaller than $k_c$, where
\be
k_c \, =  \, |\beta_1| \, = \, \frac{3}{4} \frac{g_A \lambda_1 \lambda_2}{f^2} \frac{\Omega_Q^2 Q_0^3}{m_{\chi}^2 - \Omega_Q^2} \, .
\ee
The criterion for efficiency of the resonance is
\be
\mu_k \gtrsim \frac{\Omega_Q}{2\pi},
\ee
where $\mu_k$ is the Floquet exponent describing the exponential growth of the amplitude, i.e. 
\be
C(k , t) \, \sim e^{\mu_k t} \, .
\ee
As we will show later, this condition is satisfied for parameter values of interest in the case when the vector field is a dark matter condensate.  However, in other applications, the condition may be violated.  In this case, we need to turn to the narrow-resonance channel.

In the narrow-resonance regime \(|q_\lambda|\ll 1\), the first instability band lies near
\begin{equation}
A_k \simeq 1
\qquad \Longleftrightarrow \qquad
k \simeq \frac{\Omega_Q}{2}.
\label{eq:first_band}
\end{equation}
Its width is approximately
\begin{equation}
\left|\frac{4k^2}{\Omega_Q^2}-1\right| < |q_\lambda|,
\label{eq:band_width_A}
\end{equation}
or equivalently
\begin{equation}
\left|k - \frac{\Omega_Q}{2}\right| \lesssim \frac{|\beta_1|}{4}.
\label{eq:band_width_k}
\end{equation}
At the center of the first band, the physical-time growth rate is
\begin{equation}
\mu_k \simeq \frac{|q_\lambda|\Omega_Q}{4} = \frac{k|\beta_1|}{2\Omega_Q}.
\label{eq:mu_general}
\end{equation}
Evaluated at \(k\simeq \Omega_Q/2\), this gives
\begin{equation}
\mu_k \simeq
\frac{3|g_A\lambda_1\lambda_2|}{16f^2}
\frac{\Omega_Q^2Q_0^3}{|m_\chi^2-\Omega_Q^2|}.
\label{eq:mu_center}
\end{equation}
Note that the indirect resonance is enhanced when
\begin{equation}
\Omega_Q \sim m_\chi
\qquad \text{and} \qquad
k \sim \frac{\Omega_Q}{2}.
\label{eq:double_condition}
\end{equation}
This is the clearest two-stage condition: first the scalar responds efficiently to the oscillating \(SU(2)\) condensate, then the \(U(1)\) modes resonate because they see a periodic \(\dot\chi\).

Additional bands are generated by higher odd harmonics. In the single-cosine approximation \eqref{eq:Q_cosine}, the third harmonic in \eqref{eq:Clambda_13} produces a band near
\begin{equation}
k \simeq \frac{3\Omega_Q}{2},
\end{equation}
with strength controlled by \(\beta_3\).

\section{Chirality: an Important Subtlety}

The direct axion-gauge coupling \(\chi X\tilde X\) is helicity sensitive, but the indirect resonance generated by a purely oscillatory \(Q(t)\) is not automatically chiral. Indeed, in the periodic zero-bias limit, the modulation entering Eq.~\eqref{eq:Hill_general} contains only odd harmonics, so it changes sign under a half-period shift:
\ba
\sum_{r=0}^{\infty}&\beta_{2r+1}&\cos\!\big[(2r+1)\Omega_Q(t+T_Q/2)\big] \nonumber \\
&=& -\sum_{r=0}^{\infty}\beta_{2r+1}\cos\!\big[(2r+1)\Omega_Q t\big].
\ea
Therefore, the \(+\) and \(-\) helicity equations are mapped into each other by
\begin{equation}
\lambda \to -\lambda,
\qquad
t \to t+\frac{T_Q}{2}.
\end{equation}
As a result, the Floquet exponents are identical in the perfectly symmetric limit. A net helicity asymmetry requires an additional bias, for example
\begin{enumerate}
\item a nonzero homogeneous drift \(\dot\chi_{\rm dm}\neq 0\),
\item damping or detuning that breaks the half-period symmetry,
\item backreaction effects that make the evolution nonlinear and helicity dependent.
\end{enumerate}
So the indirect \(SU(2)\)-driven effect is generically a \emph{resonant amplification} mechanism, but it is not a fundamentally \emph{chiral} one unless another ingredient is added.

There are two general applications of the formalism. The first is to a gravitationally bound halo, while the second is to an expanding cosmological patch of space. In the latter case, the expansion of space leads to the damping of the fields, which breaks the half-period symmetry mentioned above. Thus, in this case we expect a net helicity of the induced $U(1)$ vector fields.

\section{Validity of the approximation}

The analytic treatment above is based on several assumptions:
\begin{enumerate}
\item \textbf{Local Minkowski patch.} The resonance time scale \(\mu_k^{-1}\) and oscillation period \(\Omega_Q^{-1}\) are short compared to the dynamical time scale of the halo (or the expansion time if we are considering an expanding patch of space).
\item \textbf{Weak backreaction.} The produced Abelian modes do not yet significantly modify \(\chi\) through \(\langle\mathbf E_X\cdot\mathbf B_X\rangle\), and the scalar does not yet significantly distort the leading \(Q\)-oscillation.
\item \textbf{Homogeneous condensates.} Spatial gradients of \(Q\) and \(\chi\) are neglected over the local patch of interest.
\item \textbf{First-harmonic truncation.} The Mathieu reduction is accurate only if the first harmonic dominates. Otherwise, the full Hill equation \eqref{eq:Hill_general} should be used.
\end{enumerate}
Given these assumptions, the analytic picture is robust: the non-Abelian condensate generates a periodic scalar response, and the scalar response parametrically amplifies the Abelian modes.

\section{Application to Vector Dark Matter}

The formalism described in the previous sections is rather general. Now we will specialize to the case when the $SU(2)$ condensate is assumed to initially constitute the bulk of the dark matter. In the minimal quartic regime considered here, the $SU(2)$ condensate redshifts as radiation at early times; a transition to matter-like behavior would require additional structure in the gauge sector.

In this case, the energy density in $Q$ equals the radiation energy density at the time of equal matter and radiation when the temperature is $T_{eq}$, and we thus have 
\be \label{DMvalue}
g_A^2 Q(T_{eq})^4 \, \sim \, g_*(T_{eq}) T_{eq}^4 \, ,
\ee
where $g_*(T_{eq})$ is the number of spin degrees of freedom in the radiation back at $T_{eq}$.  Since the energy density in $Q$ scales as radiation at early times, we have
\be \label{scaling}
Q(T) \, = \, \frac{T}{T_{eq}} Q(T_{eq}) \quad {\rm{for}} \quad T > T_{eq} \, .
\ee

We will assume that neither $Q$ nor $\chi$ couples to the Standard Model sector. Hence, there are no plasma effects to consider and the transfer of energy from $Q$ to $\chi$ can start as soon as $Q$ begins to oscillate. This happens when the Hubble friction becomes negligible compared to $V^{\prime}(Q)$ (the prime indicating a derivative with respect to $Q$), i.e. when
\be
H^2(T) Q(T) \, = \, g_A^2 Q(T)^3 \, .
\ee
Making use of (\ref{DMvalue}), (\ref{scaling}) and the Friedmann equation to evaluate $H$, we obtain for the temperature $T = T_0$ of the onset of oscillation and the value $Q = Q_0$ of the $SU(2)$ condensate at that time,
\ba
T_0 \, &\sim& \, g_A^{1/2} g_*(T_{eq})^{1/4} g_*(T_0)^{-1/2} m_{pl} , \nonumber \\
Q_0 \, &\sim& \, \bigl( \frac{g_*(T_{eq})}{g_*(T_0)} \bigr)^{1/2} m_{pl} \, ,
\ea
where $m_{pl}$ is the Planck mass. Note that for $g_A \ll 1$, we are in the regime where effective field theory can be trusted.

We now specialize more: we assume that both the $SU(2)$ condensate and the $\chi$ condensate correspond to ultralight dark matter, whose mass we parametrize in units of $10^{-20} {\rm{eV}}$. The ``effective mass'' of the $SU(2)$ condensate, which can be defined in terms of the curvature of the potential when the field takes the value $Q(T_{eq})$, is
\be
m_A \, \equiv \, m_{20} 10^{-20} {\rm{eV}} \, \sim \, g_A Q_{T_{eq}} \, .
\ee
This leads to the result (using $T_{eq} \, \sim \, 1 {\rm{eV}}$)
 \be \label{gvalue}
g_A \, \sim \, m_{20}^2 10^{-40} g_*(T_{eq})^{-1/2} \, .
 \ee
Similarly,
\be
m_{\chi} \, \equiv \, m_{\chi, 20} 10^{-20} {\rm{eV}} \, .
\ee

We next study whether the tachyonic resonance channel is open. If $m_{20}$ and $m_{\chi, 20}$ are of $\mathcal{O}(1)$, then it follows immediately that $\Omega_Q $ and $m_{\chi}$ are of the same order of magnitude. In this case, the Floquet exponent at the large $k$ end of the tachyonic instability band is (in the single harmonic approximation) 
\be
\mu_k \, \sim \, \beta_1 \, \sim \, g_A \lambda_1 \lambda_2 f^{-2} Q_0^3, 
\ee
and
\be
\frac{\mu_k}{\Omega_Q} \, \sim \, \lambda_1 \lambda_2 \bigg( \frac{Q_0}{f} \bigg)^2 \, ,
\ee
where here, in contrast to earlier sections, $Q_0$ is the amplitude of $Q$ when plasma effects for the $U(1)$ field become negligible (as argued in \cite{BFJ} and \cite{Nirmalya}, this is the time of recombination, but see \cite{Sharma} for a different point of view). Thus,  if the coupling constants $\lambda_1$ and $\lambda_2$ are of $\mathcal{O}(1)$,  then the tachyonic resonance channel is open, provided that
\be \label{ineq}
f \, \ll \, g_A^{-1/2} T_{eq}, 
\ee
which, using (\ref{gvalue}) is about $f = m_{20}^{-1} 10^{11} {\rm{GeV}}$. (Here and in the following we are setting the factors of $g_*$ equal to one.)  If (\ref{ineq}) is violated, then the tachyonic channel is closed.

Assuming for the moment that the tachyonic channel is open, then most of the energy transferred during the tachyonic resonance goes into modes at the upper end of the tachyonic instability window,  i.e. $k \sim \beta_1$. Inserting the parameter values discussed above, we find for the corresponding comoving wavelength $\lambda_c$
\be \label{length1}
\lambda_c \, \sim \, \beta_1^{-1} \, \sim \, m_{20} (\lambda_1 \lambda_2)^{-1} 
\bigg( \frac{f}{10^{11} {\rm{GeV}}} \bigg)^2 10^{-9} {\rm{Mpc}} \, .
\ee
We can then infer the magnetic field strength which is induced on this scale. If a fraction ${\cal{F}}$ of the dark matter energy density at $T_{eq}$ is transferred to radiation, then the resulting magnetic field is of the order
\be \label{Bfield}
B \sim \sqrt{\mathcal{F}}\ \text{Gauss}.
\ee
The factor ${\cal{F}}$ is the product of two factors: first the fraction ${\cal{F}}_1$ of the $SU(2)$ condensate field which flows into the $\chi$ condensate, evaluated at $T_{eq}$, and the second is the fraction ${\cal{F}}_2$ of the $\chi$ energy which is transferred to electromagnetism during the resonance stage. The first factor is determined by (\ref{frac1}) and (\ref{frac2}), taking into account the factor that the enegies in $\chi$ and $Q$ redshift differently. The second factor is determined by when backreaction shuts off the resonance.

Now let us move on to the case when the tachyonic channel is closed, i.e. when (\ref{ineq}) is violated.  We work in the single harmonic approximation. In this case, the first and leading resonance band is centered at
\be
k \, \sim \, g_A Q_0 \, ,
\ee
where again $Q_0$ is the amplitude of oscillation at the time of recombination, which is of the order of $g_A^{-1/2} T_{eq}$. This corresponds to a wavelength of
\be \label{length2}
\lambda \, \sim \, m_{20}^{-1} 10^{-9} {\rm{Mpc}} \, .
\ee
Since the Floquet exponent is parametrically much larger than the Hubble expansion rate at the time of recombination, the parametric resonance instability is efficient and can transform a large fraction of the energy in the $\chi$ condensate into magnetic fields. Thus, the resulting magnetic field strength is given by the same expression as in (\ref{Bfield}), except that the backreaction effects which determine the factor ${\cal{F}}$ will be different.

\section{Summary}

The indirect effect of the oscillating \(SU(2)\) condensate on the Abelian field can be summarized as follows:
\begin{enumerate}

\item The isotropic non-Abelian condensate \(Q(t)\) obeys a nonlinear oscillator equation with periodic solutions.

\item Through the coupling \(\chi F\tilde F\), the quantity \(Q^2\dot Q\) acts as a periodic source for  \(\chi(t)\), the homogeneous scalar condensate.

\item The induced \(\dot\chi(t)\) enters the \(U(1)\) mode equation as a periodic helicity-dependent frequency shift.

\item The resulting Abelian mode equation is a Hill equation, and in the leading-harmonic regime it becomes a Mathieu equation with resonance bands near \(2k\simeq (2r+1)\Omega_Q\).

\item The strongest enhancement occurs when the scalar is itself close to forced resonance, \(m_\chi \simeq (2r+1)\Omega_Q\), producing a two-stage cascade \(Q \to \chi \to U(1)\).

\item Applied to an ultralight vector dark matter model,  the scenario is able to generate large amplitude magnetic fields immediately after the time of recombination on a length scale given by (\ref{length1}) or (\ref{length2}), depending on whether the tachyonic instability channel is open or not.

\end{enumerate}

The main analytic lesson is therefore simple: an oscillating \(SU(2)\) condensate can indirectly trigger parametric resonance of a \(U(1)\) field, even though the two gauge sectors do not couple directly. The scalar \(\chi\) acts as the mediator, and the entire mechanism is encoded in the periodic structure of \(\dot\chi(t)\).

\acknowledgments

\noindent 
The research at McGill is supported in part by funds from NSERC and from the Canada Research Chair program.   TD is supported in part by a Trottier Space Institute postdoctoral fellowship. VK would like to acknowledge the McGill University Physics Department for hospitality.   



\end{document}